\documentclass[10pt,conference]{IEEEtran}
\newtheorem{lemma}{Lemma}

\newcommand{\bS}{{\mathbf{S}}}

\newcommand{\bU}{{\mathbf{U}}}

\newcommand{\bv}{{\mathbf{v}}}

\newcommand{\cV}{{\mathcal{V}}}

\newcommand{\bX}{{\mathbf{X}}}

\newcommand{\bY}{{\mathbf{Y}}}

\newcommand{\modulo}{\mbox{\,mod\,}}

\newcommand{\bN}{\mathbf{N}}

\begin{document}

\title{A Simple Extension of the $\modulo$-$\Lambda$ Transformation
%to Multiple-Access Channels
}
\author{Uri Erez and Ram Zamir}
\maketitle

\begin{abstract}
A simple lemma is derived that allows to transform a general scalar (non-Gaussian, non-additive) continuous-alphabet channel as well as a general multiple-access channel into a modulo-additive noise channel. While in general the transformation is information lossy, it allows to leverage
linear coding techniques and capacity results derived for networks comprised of additive Gaussian nodes to more general networks.
\end{abstract}

\section{Introduction}
There has been considerable progress in recent years in deriving coding schemes and capacity results for scalar channels as
well as networks with AWGN noise, e.g., \cite{EZ04,NG07,PKEZ07}. A key ingredient in these results is the use of linear/lattice codes which are
particularly well suited for additive noise channels. In this note we derive a lemma that allows to leverage
these techniques to networks which have general multiple-access (MAC) channels nodes.  Specifically, we provide a transformation that
converts a general (neither Gaussian nor additive) continuous-amplitude MAC channel, into an modulo-additive one, at the price of some information loss.
It is envisioned that in certain cases, the benefits of lattice coding will outweigh the incurred loss in mutual information. More specifically,
for any network of Gaussian channels for which lattice encoding schemes are exploited to derive an achievable rate region, using the derived lemma, one may
obtain an analogous inner bound on the rate region for a network with the same topology but with general channels.
The transformation is a simple extension of the $\modulo$-$\Lambda$ transformation derived for scalar channels as given
in \cite{EZ04} and specifically the version given by Forney in \cite{ForneyAllerton}.

\section{$\modulo$-$\Lambda$ transformation}
For simplicity we consider a two-user MAC channel. The extension to the case of more users is straightforward.
We thus assume a two-user MAC channel with inputs $X_1,X_2$
and output $Y$, where all random variables are real-valued and continuous.
We further assume that the transmitters are subject to the same power constraint
$E[X_i^2] \leq P$.
%We may transform the channel into a $\modulo$-$\Lambda$ MAC channel as we describe next.
Let $\Lambda$ be a lattice with fundamental region $\cV$ having second moment $P$. Let $\bv_i \in \cV$, $i=1,2$, be the information bearing signals
of the users. Further, let $\bU_i \sim {\rm Unif}(\cV)$, $i=1
,2$,  be dithers uniformly distributed over $\cV$.
We assume that the dither $\bU_i$ is known to transmitter $i$ as well as to the receiver\footnote{In practice, a pseudorandom sequence
will be used at both transmission ends.}.

\noindent {\bf Transmitter $i$ sends:}
\begin{eqnarray}
\bX_i=\bv_i+\bU_i \modulo \Lambda
\label{l1}
\end{eqnarray}
Note that due to the dither, $\bX_i$ is uniformly distributed over $\cV$ \cite{EZ04,ForneyAllerton},
for any value of $\bv_i$. In particular, $\bX_i$ satisfies the power constraint $P$.
Let $\bS=\bX_1+\bX_2$ and
let $\hat{\bS}=g(\bY)$ be an estimator of
$\bS$ from the output $\bY$.
Ideally, the criterion for estimation would be that of minimum noise entropy but for simplicity one could use the MMSE estimator or even a linear MMSE estimator.
Denote the estimation error by $\bN=\hat{\bS}-\bS$. The receiver processes the output as follows,

\noindent{\bf Receiver computes:}
\begin{eqnarray}
\bY' & = & \hat{\bS}-\bU_1-\bU_2 \modulo \Lambda \nonumber \\
     & = & g(\bY)-\bU_1 -\bU_2 \modulo \Lambda.
\label{l2}
\end{eqnarray}
We now observe that the induced channel from $\bv_1,\bv_2$ to $\bY'$ is indeed additive modulo $\Lambda$.
Let us add and subtract $\bS$ from the r.h.s. of (\ref{l2}) to obtain:
\begin{eqnarray*}
\bY' & = & \hat{\bS}+(\bS-\bS)  -\bU_1-\bU_2 \modulo \Lambda \\
     & = & \hat{\bS}+(\bv_1+\bU_1+\bv_2+\bU_2)  -\bS -\bU_1-\bU_2 \modulo \Lambda  \\
%     & = & \bv_1+\bv_2+(\hat{\bS}-\bS)  \modulo \Lambda  \\
     & = & \bv_1+\bv_2+{\bf N}  \modulo \Lambda,
\end{eqnarray*}
where $\bN$ is independent of the inputs.
%Note that $(\bV_1,\bV_2,\bU)\leftrightarrow(\bX_1,\bX_2) \leftrightarrow \bY$ forms a Markov chain. Since $\bX_1,\bX_2$ are
%independent of $\bV_1,\bV_2$, so is the pair $(\bS,g(\bY))$  and hence $\bN$.
We thus obtain the following lemma,
\begin{lemma}
Applying the transmission scheme (\ref{l1}),(\ref{l2}), results in a $\modulo$-$\Lambda$ additive noise channel:
\begin{eqnarray*}
\bY' & = & \bv_1+\bv_2+{\bf N}  \modulo \Lambda
\end{eqnarray*}
where $\bN=\hat{\bS}-\bS$ is independent of the inputs $\bv_1$ and $\bv_2$.

\end{lemma}
\section{Discussion}
The lemma generalizes the scalar version of \cite{EZ04} in two ways. First, it applies to
any channel, which need not be Gaussian nor additive, and thus extends results even for a scalar channel.
Second, it treats MAC channels as well.
We note that while in \cite{ForneyAllerton}
the optimal estimator turned out to be linear since the channel considered was an AWGN channel, it is apparent that
in a more general setting, as the one at hand, non-linear estimation is beneficial.
Finally, one may extend the results presented by allowing each transmitter to perform some (in general non-linear)
preprocessing of its input signal.

\end{document}